# Cross-pollination dynamics of web-based social media: An application of insect-mediated pollen transfer

Raul A. Barreto[1,2,3], Angus Flavel[2,3] and Raphael Xelot-Wilson[3]

[1] School of Economics and Political Science, The University of Adelaide

[2] Flavel Research Electronics and Manufacturing (REM) Pty. Ltd.

[3] Share Social Pty. Ltd.

## Author Note

Raul A. Barreto, https://orcid.org/0000-0001-8578-5131.

Correspondence concerning this article should be addressed to Dr. Raul Barreto, School of Economics and Political Science, The University of Adelaide, Adelaide SA, 5005 Australia, Email: raul.barreto@adelaide.edu.au.

Angus Flavel and Raul Barreto are the CEO and CFO, respectively, of both Flavel REM Pty Ltd and Share Social Pty Ltd. The latter company is the sole proprietor of the **&share** app.

This paper is the result of research conducted by Flavel REM Pty Ltd under contract by Share Social Pty Ltd as part of the latter's Australian AusIndustry Research and Development Tax Incentive (RDTI) reporting obligations.

A. Flavel conceived the presented idea. R. Barreto and R. Xelot-Wilson developed the methodology, R. Barreto developed software usage, performed the validation, and conducted the formal analysis and investigation. A. Flavel provided resources and data curation. All three authors contributed to writing the original draft, review and editing. R. Barreto provided supervision. A. Flavel provided project administration. R. Xelot-Wilson provided technical support.

# Cross-pollination dynamics of web-based social media: An application of insect-mediated pollen transfer

## Abstract

We propose a model of cross-pollination among online social media (OSM) websites, where the dynamics of user interactions mimic insect-mediated pollen transfer by pollinators. A pollinator acts as a vehicle enabling users to visit multiple social media sites—akin to visiting different plants in the same field—within a single browsing session. This approach frames geitonogamy in self-incompatible plant species as analogous to the distribution of web traffic across the social media landscape. A theoretical pollinator, allowing users to choose among social media sites multiple times per trip, drives uneven increases in web traffic across platforms, disproportionately benefiting the largest social networks while providing tangible competitive advantages to smaller OSMs. This heterogeneous landscape fosters monopolistic competition among niche platforms, incentivizing smaller sites to promote cross-pollination despite the larger relative gains to their bigger competitors. Our findings underscore the broader value of cross-platform user engagement and parasocial relationships, highlighting how cross-pollination dynamics can intensify network effects and bolster interconnectivity. Cross pollination via new pass-through apps facilitates the movement of attention, deepening and distributing engagement across multiple destinations. As pass-through apps gain traction, their disproportionate impact on traffic to social media platforms will incentivize social media platforms, large and small, to embrace cross-pollination dynamics.

**Introduction**

The internet—and, more recently, social media—has reshaped society in profound ways. Today, nearly every action we take and good we consume is directly or indirectly affected by the vast, interconnected web of information available online, enabled by seamless smartphone access and rapid social media dissemination. As this ocean of content continues to expand and becomes increasingly accessible, the depth of interconnections between users and content also grows.

In 2024, 66.2 percent of the global population accessed the internet, and 62.3 percent held social media identities, leaving a penetration gap of just 3.9 percentage points—a sharp increase from 30 percent internet and 22 percent social media penetration, with an 8 percent penetration gap, in 2012. Figure 1 illustrates that the penetration gap between internet and social media usage, which remained steady between 2012 and 2014, later widened to peak at 15 points in 2018 before gradually narrowing to current levels.

Since 2015, average daily internet usage has risen from 4.40 hours (Kemp, 2015) to 6.40 hours in 2024 (Kemp, 2024). Conversely, social media usage has slightly decreased, from 2.40 hours in 2015 to 2.23 hours in 2024, with national variations ranging from 0.7 hours in Japan to 4.3 hours in Argentina in 2015, and from 0.53 hours in Japan to 3.43 hours in Kenya in 2024. This shift may reflect the growing popularity of other content forms, such as video streaming (3.06 hours), music streaming (1.25 hours), and podcasts (0.49 hours) in 2024. These other forms represent the growing impact of parasocial relationships amongst content consumers leading to greater behavioural loyalty with their favorite content creators (See Lim, et.al., 2020) and the increased likelihood to seek out their pages on other platforms.

Social media is distinct from other forms of media: it functions as a two-sided platform that hosts user-generated content, disseminated algorithmically and enhanced through user interactions. This definition excludes streaming services, which lack a social component; discussion forums, which lack the media component; and email, lacking the platform component (Aridor et al., 2024).

**Figure 1**[1]

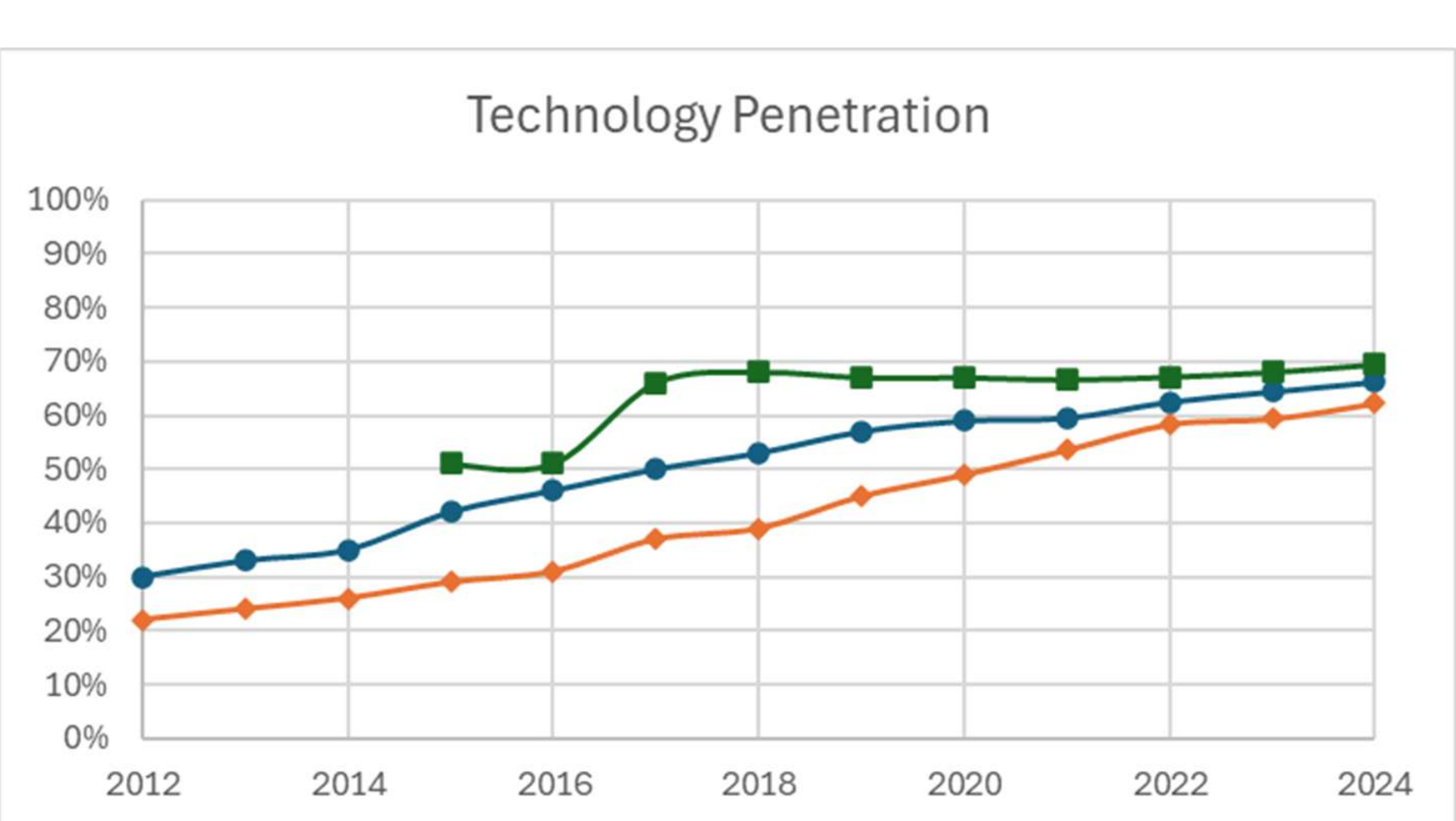

As internet and social media penetration rates converge, anecdotal evidence suggests a similar convergence between social media and other previously distinct online services. For example, playlists from audio streaming services are regularly shared on social media; online discussions flourish on video streaming platforms; hot links populate discussion forums; and email identities often serve as public online profiles. As interconnectivity grows, boundaries blur between social media and other online media services.

We suggest that the last paradigmatic shift, as defined as defined by Kuhn (1962), in internet usage occurred with the emergence of the search engine. Google's ascent in the early 2000s established a new frontier for access, structure, and discovery across the digital landscape. What followed was a series of fragmented evolutions- successive waves of social media platforms layered functionality upon one another, refining user interfaces, tightening feedback loops, and adjusting the units of content that defined their ecosystems. From microtext to vertical video, each platform iteration altered the format of media but did not revisit the architecture beneath it.

[1] See Kemp (2024, 2023, 2022, 2021, 2020, 2019, 2018, 2017, 2016, 2015, 2014, 2013, 2012).

In the years since, no single actor—corporate, individual, or institutional—has challenged the structural adequacy of the search engine itself. Nor has any platform offered a coherent unifying strategy for the coexistence of discrete and incompatible search functions embedded within social media applications. Instagram, TikTok, and X each possess isolated search bars, yet none can resolve the broader query of who someone is, where they reside online, and how their digital presence spans multiple ecosystems. Each platform governs its own walled garden, with no interoperable channel for media discovery across boundaries.

We propose a model of cross-pollination among online social media (OSM) platforms, drawing an analogy to insect-mediated pollen transfer. Following the theoretical pollination model of Di Pasquale and Jacobi (1997), we conceptualize users as bees that "pollinate" the digital ecosystem by visiting multiple social media sites within a single browsing session, or "trip." In this analogy, bees' geitonogamy and pollen export within self-incompatible plant species mirror the distribution of web traffic across the social media landscape. Here, individual webpages within OSMs represent the "pollen" that attracts users.

Our model illustrates, across both deterministic and stochastic scenarios, that a "pollinator" application—allowing users to select and navigate between multiple social media sites per trip—can drive increased web traffic across all platforms, with the largest social media networks capturing most of this growth. However, despite the larger sites' advantage, user heterogeneity also fosters monopolistic competition among smaller social media networks, motivating even the smallest platforms to support cross-pollination despite the comparatively greater benefit to larger competitors.

We observe that that while platforms are siloed and users are multi-home, search is fragmented. Hence, we ask, what governs cross-platform attention allocation? We model browsing as a stochastic cross platform Markov process with endogenous stopping to define a closed-form expression for attention allocation. We find that cross-platform navigation amplifies concentration endogenously and explicitly, concluded cross-platform navigation amplifies dominance of large platforms.

Cross-pollination among social media platforms is currently happening organically via apps such as LinkTree, Link.Me and &Share. They introduce a systemic rethinking of navigation, contact discovery, media search, and platform interoperation by addressing the inefficiencies of access and continuity. Pollinator apps do not create content, rather they facilitate the movement of attention by deepening and distributing engagement across multiple destinations. The process introduces a positive-sum effect across the platform ecosystem, increasing traffic, enhancing advertising yield, and exposing users to content that would otherwise remain siloed. Importantly, it does so without requiring users to abandon their preferred environments, but instead by enabling them to traverse those environments with cohesion and intent. We suggest that the expanding diversity of online content, paired with a stable user distribution, will continue driving the convergence between internet and social media penetration via increasingly pronounced cross-pollination.

Drawing insights from nature to better understand complex societal relationships is a well-established approach. For example, in the information technology field, matching algorithms inspired by natural behaviours include those based on krill movement (Gandomi and Alavi, 2012), cuckoo breeding behaviour (Nguyen et al., 2015), firefly attraction (dos Santos Coelho et al., 2013), and flower pollination (Yang, 2012). In economics, such applications are common (see Bourgeois-Gironde et al., 2021; Dener et al., 2016; Addessi et al., 2019; de Waal, 2021), especially in neuroeconomics, where researchers integrate psychology and economic methodologies with neuroscience to investigate animal decision-making processes (Hayden et al., 2010; Calvert et al., 2011; de Visser et al., 2011). Similarly, researchers like Barreto (2018, 2024) and Alm and Barreto (2024) have applied fluid dynamics principles from engineering to model dynamic economic systems.

The idea of social media cross-pollination first appeared in the literature with Jain et al. (2013), who found that the most significant benefit of cross-pollination is increased traffic and user engagement on the platform where the pollinator lands. However, to our knowledge, no previous studies have examined the systemic behavioural characteristics of insect-like cross-pollination as a metaphor for user interactions across social media sites.

## The Model

There currently exist several social media agglomeration apps that gather all the social media handles of any person or organization in one place.[2] These pass-through apps are the metaphorical bees, or pollinators. The internet user on the pollinator app is drawn to the OSM webpages—analogous to pollen. The systemic sequence of path-dependent landings on pages across the social media landscape by any user represents the trip that a bee takes, moving from flower to flower, plant to plant, and field to field in search of pollen.

Suppose an internet user finds themselves on a social media page of a popular figure or content creator, defined in figure 2 by the oval shape. A click to the pollinator app, defined by the '&' symbol, begins the trip. The metaphorical field contains the various species flowers, represented by the choice of OSMs, as defined in figure 2 by the different colours where the same popular figure as a presence.

---

[2] Examples include Link.Tree, Link.Me and &Share. They range in motivations and target markets but provide the same fundamental service of consolidating user handles into a single account or "stream" as in the case of &Share.

The pollinator app transports the user from their initial landing to any one of the personality's other OSM pages. The user then engages with the OSM beyond the popular figure's presence there, possibly extending the trip to other fields by visiting different pages within the OSM, or returning to the pollinator app, extending the same trip by visiting another of the personality's OSM pages—a different flower within the same field.

We formalize the problem by assuming state space $S$ of $m$ personalities across $n$ OSMs, $S = \{(m, n)\}$. The user lands on page $(m,n)$ and spends some time there, $\tau_{mn} > 0$. Conditional on continuing the session, the user transitions to another page according to the transition matrix, $Q \in \mathbb{R}^{|S| \times |S|}$, with entries defined in equation (1.2).

$$S = \{(m, n) : \text{personality } m \text{ has a landing page on OSM } n\} \tag{1.1}$$

$$q_{(m,n),(m',n')} = \Pr\left((m', n') \, \text{next} \,\middle|\, (m, n), \text{session continues}\right) \tag{1.2}$$

Each element of $Q$ is defined as follows.

$$q_{ij} = \frac{A_j^{\theta}}{\sum_k A_k^{\theta}} \tag{1.3}$$

$A_j$ is the attraction to the OSM- content stock, user base, etc.- and $\theta > 0$ is the attention sensitivity.

We define expected total browsing time $\overline{T}(\sigma)$ in equation (1.4), where $\sigma$ is the initial distribution over $S$ after the user clicks the pollinator from the original landing page $(m,n)$, $\tau$ is the average time spent on visited page.

**Figure 2**

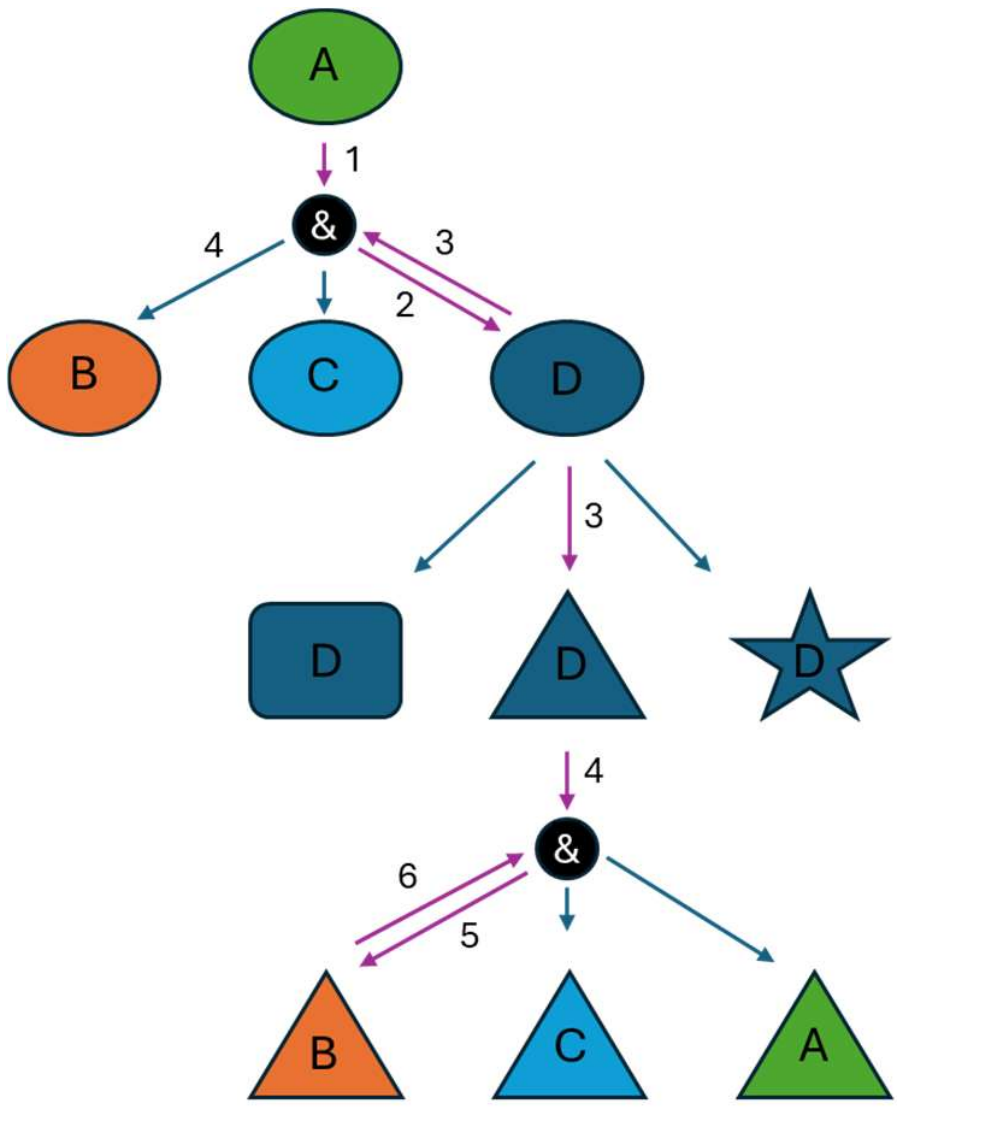

$$\overline{T}(\sigma) = \tau\left(\sum_{k=0}^{\infty} A_k Q^k\right)\sigma \tag{1.4}$$

$A_k$, defined in equation (1.5), is the stage weighted probability the session has survived to this point, where $\alpha_k \in [0,1]$ is the probability that the trip continues after the *k*-th visited state.

$$A_k = \prod_{r=1}^{k} \alpha_r \quad ,(k \geq 1) \tag{1.5}$$

We define the expected number of stops at any transient state by the fundamental matrix, *N*.[3]

$$N = \sum_{k=1}^{\infty} A_k Q^k \leq (1-Q)^{-1} \tag{1.6}$$

Convergence follows from the spectral radius of *Q* being less than one. The (*i*,*j*)th element of *N* gives the expected number of visits to state *j* starting from state *i*, prior to termination of the browsing session. We can adjust equation (1.6) for continuation probabilities such that equation (1.7) expresses the expected number of stops from each starting state.[4]

[3] Equation (1.6) is convergent due to the spectral norm of *Q* being inferior to 1

$$N_\alpha = \sum_{k=0}^{\infty} A_k Q^k = I + \sum_{k=1}^{\infty} \left( \prod_{r=1}^{k} \alpha_r \right) Q^k \quad (1.7)$$

Equation (1.7) aggregates all possible browsing sequences, weighting each *k*-step transition by the probability that the session persists to that stage. Accordingly, total expected browsing time compacts to $\overline{T}(\sigma) = \tau N_\alpha \sigma$.

The increase in direct traffic from the pollinator is a function of the ability of personalities on the OSM to capture and hold the attention of the user, $\alpha_k$. In other words, the greater the depth and scope of a social media userbase, the greater the direct benefit from direct cross-pollination. The benefit to the largest OSM is further increased indirectly as the user interacts with other personalities, who attract online attention heterogeneously. The OSM with the greatest depth of personalities that commands the longest attention from its users per visit will reap greatest indirect benefits from the pollinator.

*Theorem 1*: Platform dominance under cross-platform browsing = the pollinator benefits large platforms more than smaller ones.

Suppose there $S_n \subset S$ states belonging to platform *n*. Equation (1.8) represents the total expected time captured by the platform *n* is the portion of $\overline{T}(\sigma)$ attributable states on platform *n*.

$$G_n(\sigma) = \sum_{i \in S_n} \tau \left[ \left( \sum_{k=0}^{\infty} A_k Q^k \right) \sigma \right]_i \quad (1.8)$$

[4] Equation (1.7) is bounded above by $(1-Q)^{-1}$.

Let $s_n$ measure platform *n*'s size, interpreted as breadth of reachable personalities, traffic, or attractiveness. Assuming (1) entry monotonicity, then bigger platforms receive more initial pollinator traffic and for every initial distribution $\sigma$, the total initial mass assigned to platform *n*, $\sum_n(\sigma) \equiv \sum_{i \in S_n} \sigma_i$ is weakly increasing in $s_n$. Assuming (2) transition monotonicity, then bigger platforms attract more follow-on transactions and for every state $j \in S$, the total probability of next-step movement into platform $n$, $M_n \equiv \sum_{i \in S_n} q_{ij}$ is weakly increasing in $s_n$.

Assuming (3) retention monotonicity, conditional on larger platforms receiving larger share of subsequent visit, bigger platforms better retain and monetize than their smaller counterparts, such that $\overline{\tau}_n = \frac{1}{|S_n|} \sum_{i \in S_n} \tau_i$ is weakly increasing in $s_n$. Hence for any two platforms *m* and *n*, if $s_n > s_m$, if follows that $G_n(\sigma) \geq G_m(\sigma)$ for every initial distribution of $\sigma$ and if either monotonicity (1) through (3) strictly holds, then $G_n(\sigma) > G_m(\sigma)$. The series in equation (1.8) assigns more expected visits, and hence more expected total time, to larger platforms. More simply, $\frac{\partial G_n}{\partial s_n} > \frac{\partial G_m}{\partial s_m}$ for all $s_n > s_m$. In words, larger platforms capture weakly more expected total attention than smaller platforms under cross-platform browsing.[5]

The cross-pollination's impact on the social media landscape creates a strictly positive-sum gain. While traffic rises across all platforms, the depth of personalities and their prominence on more heavily trafficked social media channels skew the distribution of these gains in favour of larger OSMs at the expense of smaller ones.

Suppose the initial landing page is on a relatively small OSM serving a niche audience. Once engaged with the pollinator, the user will likely navigate to one of the personality's other OSM pages. Statistically, the user will tend to land on the personality's page within the largest social media platform with the highest traffic.

[5] See Proof of theorem 1 in Appendix 1.

Large OSM sites have a clear advantage in capturing users from the pollinator. Nevertheless, smaller sites have a strong incentive to participate in cross-pollination, even though it may benefit larger OSMs relatively more, as their user base is often defined by a specific niche. For smaller OSMs, joining cross-pollination helps grow their niches by tapping into the vast number of users who pass through the pollinator. Ultimately, every small OSM is compelled to participate in cross-pollination because their direct competitors—other small OSMs—are doing so as well.

The baseline model describes the probabilistic effect of a user being confronted with a limited choice among OSMs. The choice is curated based on which OSMs the personality has a presence on. Analytically, all social media platforms with a presence on the pollinator benefit from increased traffic.

Cross-platform navigation creates a reinforcement process. Large platforms benefit from cross pollinations due to higher entry probability, higher retention and higher re-exposure. That benefits link directly to network effects, preferential attachment and attention markets.

Imagine the personalities within the pollinator, in addition to links to their other OSM pages, also maintain a curated pool of content available to the user. The user is thus exposed to an expanded choice that includes the personality's OSM pages as well as a pool of curated hyperlinks to places across the social media landscape that the personality finds interesting. This is analogous to the user being allowed to sample the personality's own pollen sack. Figure 3 illustrates this additional choice to visit the personality's curated content pool across OSM pages.

**Figure 3**

From the perspective of the OSMs, the extra pool option has no net impact on the views received. Remember, the probability of reaching any OSM is a function of that OSM's traffic. Therefore, the content pool of curated links by any given personality will mirror the traffic of the sites to which they lead. In aggregate, there is no change in the additional views to the OSM resulting from the inclusion of the extra option in personality content pools, as the probabilities within any given pool reflect the underlying probabilities among OSMs.

## Discussion

Cross-pollination in OSMs is already occurring organically. As expected, the largest social media networks command the greatest traffic. While the average percentage of OSM account holders who use other social media platforms remains relatively stable, the percentage of random users on an OSM who also hold accounts on other OSMs is declining as the service's popularity grows. Figure 4 suggests structural differences between visitors to the four most popular OSMs and visitors to other platforms. Visitors to the less established OSMs are necessarily less likely to have accounts on the large OSMs. In other words, cross-pollination across the OSMs is occurring and appears positively correlated with OSM traffic.

**Figure 4**

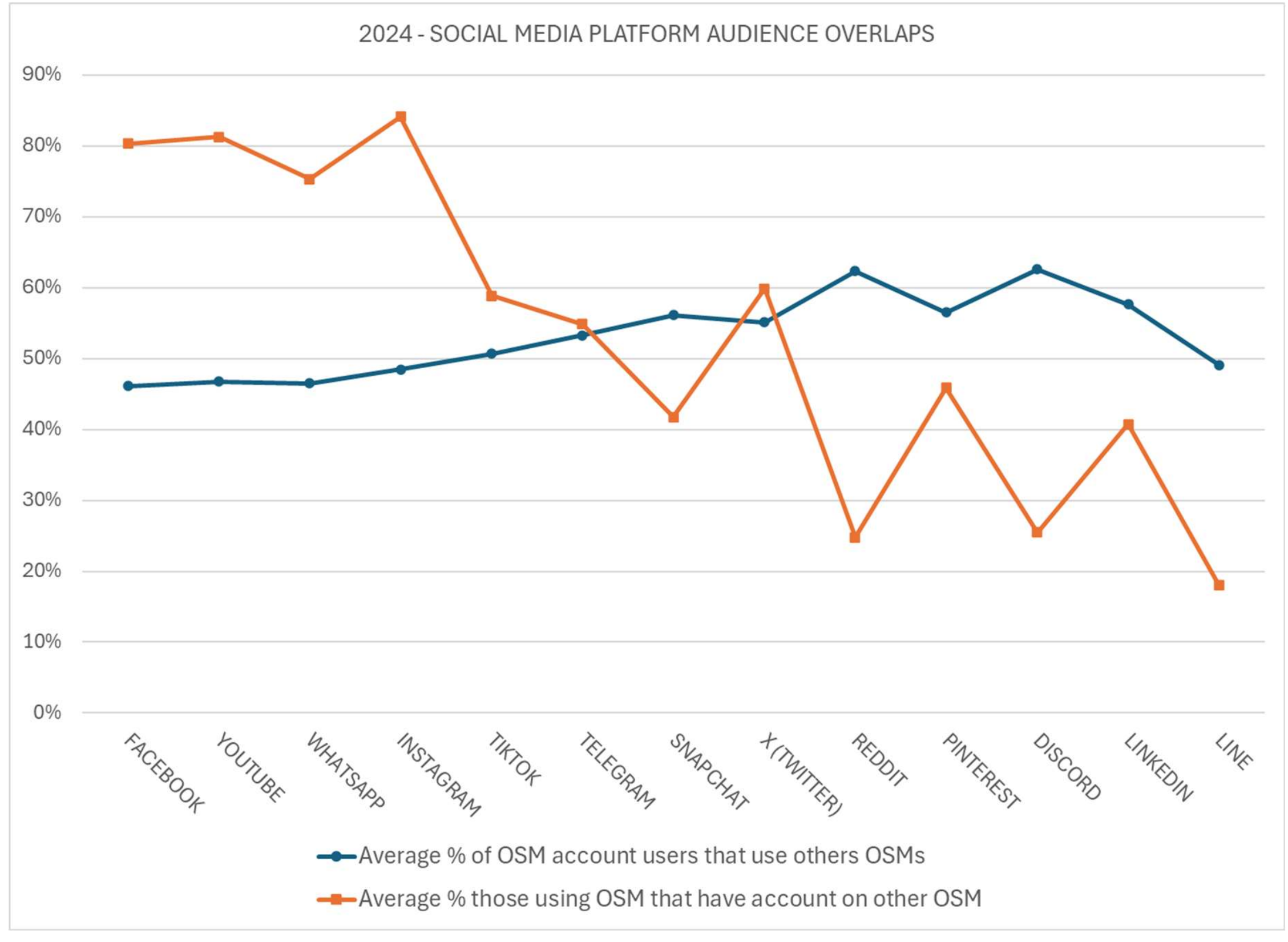

In addition to traffic volume, there is the time each user spends on social media. Extrapolating from global Android usage data between 01 July 2023 and 30 September 2023, we can estimate the daily time spent on each social media platform by a representative user. Comparing Figures 5 and 6 illustrates the difference between traffic volume and time spent on each platform. For example, TikTok has lower volume but higher engagement, while engagement on smaller platforms such as Snapchat and Pinterest appear to be less than their traffic volume would suggest. Taken together, smaller OSMs struggle to reach the mass audiences of larger OSMs as the user bases of those larger platforms become more homogeneous, gradually reducing their ability to attract users from either tail of the distribution.

**Figure 5**

Minutes per day spent on OSM by representative user
July 2023 to September 2023

FACEBOOK & FB Messenger 22 min
YOUTUBE 38 min
WHATSAPP Messenger 17 min
INSTAGRAM 15 min
TIKTOK 34 min
3
3
4
1
TELEGRAM SNAPCHAT X (TWITTER) PINTEREST

So. www.data.ai

Figure 6 illustrates the frequency with which users log on to OSMs during the same period covered by Figure 5. While users log on to WhatsApp almost 6 days a week, they spend only 17 minutes there. TikTok and YouTube show a similar pattern; their users log on at least 4 days a week and spend the most time on these platforms. Interestingly, Instagram has the same log-on frequency, over 4 days, but users spend less time on the platform.

Considering both the frequency and time spent on each platform, we can compare each platform's depth of engagement in figure 7. To account for diminishing returns in attention spans, we consider the natural log of seconds per day on the OSM. Analytically, this approach implies that the greatest relative value of time is within the first minute of arriving on any given social media platform.

Our methodology necessarily weighs platforms like X (Twitter), Snapchat, Pinterest, and Telegram, which have less meaningful engagement but higher frequency, more heavily.

**Figure 6**

Frequency of use:
Days per week users logon

FACEBOOK 4.5
YOUTUBE 4.5
WHATSAPP 5.8
INSTAGRAM 4.3
TIKTOK 4.3
TLEGRAM 2.7
SNAPCHAT 2.8
X(TWITTER) 3.0
PINTEREST 1.9
LINKIN 1.8
FB MESSENGER 3.0

▪ FACEBOOK ▪ YOUTUBE ▪ WHATSAPP ▪ INSTAGRAM ▪ TIKTOK ▪ TELEGRAM ▪ SNAPCHAT ▪ X (TWITTER) ▪ PINTEREST ▪ LINKEDIN ▪ FB MESSENGER

So. www.data.ai

We differentiate between traffic volume—the determinant of clicks-per-thousand-impressions (CPM) advertising revenue—and depth of engagement—the determinant of cost-per-click (CPC) advertising revenue.

The gross implication of cross-pollination among online social media platforms by a dedicated pollinator app is increased traffic, higher engagement, and additional advertising revenue from this increased activity. Table 1 provides advertising revenue indices to compare the revenue effects of cross-pollination of content across OSMs.

At any given time, the probability of engagement by the representative user with any respective OSM is determined by the percent of total social media time spent per OSM per day, denoted as %MT. Assume the OSM charges advertisers a $2 cost per click (CPC), with an expected click-through rate determined by the probability of landing, %MT, multiplied by

**Figure 7**

Depth of engagement:
frequency * LN(minutes/day)

FACEBOOK 31.6
YOUTUBE 34.4
WHATSAPP 40.3
INSTAGRAM 29.3
TIKTOK 32.8
TELEGRAM 14.7
SNAPCHAT 14.6
X(TWITTER) 16.8
PINTEREST 18.2
LINKIN 7.2
FB MESSENGER 46.5

So. www.data.ai

the depth of engagement, DEPTH. The CPC weekly revenue index measures the traffic and engagement-weighted revenue per OSM from the representative user. Between July 1, 2023, and September 30, 2023, the representative user contributed to CPC index revenue as follows: $10.39 to Facebook, $18.34 to YouTube, $9.64 to WhatsApp, etc.

Further assume that each OSM charges advertisers a $7 cost per one thousand impressions (CPM). We calculate the CPM revenue index for the period as the frequency of visits per week multiplied by the time spent by the representative agent on the OSM. Note that OSMs with greater frequency but lower engagement are reflected in the indices. Between July 1, 2023, and September 30, 2023, the representative user contributed to CPM index revenue as follows: $7.11 to Facebook, $11.63 to YouTube, $6.82 to WhatsApp, etc..

**Table 1**

| Online Social Media Implied Advertising Revenue: 1 July 2023 - 30 September 2023 | | | | | | | | | |
|---|---|---|---|---|---|---|---|---|---|
| | TIME | LN(TIME) | %MT | FREQ | DEPTH | CWRI | MWRI | ΔCWRI | ΔMWRI |
| | Minutes spent per OSM per day | Natural Log of time in seconds spent per OSM per day | Percent of total Social Media time spent per OSM per day | Frequency of visits per week per OSM | Depth of engagement = Freq. x Ln(time) | CPC Weekly Index Revenue = $2 x %MT x DEPTH | CPM Weekly Index Revenue = $7 x FREQ x TIME / 100 | CPC index revenue per pollination per OSM = %MT x CWRI | CPM index revenue per pollination per OSM = %MT x MWRI |
| FACEBOOK & FB Messenger | 22.6 | 7.21 | 16.20% | 4.5 | 32.37 | $10.49 | $7.11 | $1.699 | $1.152 |
| YOUTUBE | 37.3 | 7.71 | 26.67% | 4.5 | 34.39 | $18.34 | $11.63 | $4.893 | $3.102 |
| WHATSAPP | 16.7 | 6.91 | 11.98% | 5.8 | 40.25 | $9.64 | $6.82 | $1.155 | $0.817 |
| INSTAGRAM | 15.5 | 6.83 | 11.09% | 4.3 | 29.47 | $6.54 | $4.68 | $0.725 | $0.518 |
| TIKTOK | 33.3 | 7.60 | 23.81% | 4.3 | 32.82 | $15.63 | $10.06 | $3.722 | $2.395 |
| TELEGRAM | 3.7 | 5.39 | 2.63% | 2.7 | 14.42 | $0.76 | $0.69 | $0.020 | $0.018 |
| SNAPCHAT | 3.5 | 5.34 | 2.49% | 2.8 | 15.06 | $0.75 | $0.69 | $0.019 | $0.017 |
| X (TWITTER) | 4.6 | 5.61 | 3.27% | 3.0 | 16.58 | $1.08 | $0.94 | $0.035 | $0.031 |
| PINTEREST | 1.8 | 4.67 | 1.27% | 1.9 | 8.89 | $0.23 | $0.24 | $0.003 | $0.003 |
| LINKEDIN | 0.8 | 3.91 | 0.60% | 1.8 | 6.90 | $0.08 | $0.10 | $0.000 | $0.001 |

Before the introduction of the pollinator app, assume the OSMs' index revenue stream is determined by CWRI and MWRI. Inclusion of the pollinator improves revenue by adding an extra visit from the original landing on the personality's page to another of the personality's pages on a different OSM. If we assume the probability of the second landing is equal to the percentage of social media time spent per OSM per day, %MT, then the additional revenue per level of cross-pollination is %MT multiplied by either index revenue.

Revenue from pollinator traffic to any OSM is strictly greater than or equal to the original revenue from the landing. Users landing on a personality's page will either search for another personality within the same OSM, allocating time based on the first element of Equation [1], or exit the OSM to the pollinator. Since initial user traffic is distributed across $N$ OSMs while secondary pollinator traffic is distributed across $(N - 1)$ OSMs, the time added to the trip for every visit to the pollinator app, as defined by the second element of Equation [1], is strictly positive and reflects a continuous search theme.

There are two fundamental points that merit discussion regarding the nature of the increased traffic from the pollinator. The first issue is the quality of the traffic. Is the quality of traffic reaching the OSMs directly from the personality pools different from other types of traffic? The second issue pertains to the users, whom we have assumed to be homogeneous until now. What if we instead account for user heterogeneity?

The pollinator results in increased traffic across the social media landscape. It is important to note that traffic can enter the OSM from the pollinator via two distinct user preference structures.

If the user moves from the landing page of a personality on OSM A to another landing page of the same personality on OSM B, they will necessarily investigate the new landing page and possibly click on one of that personality's pieces of content. The user will have some level of engagement with the personality's content on OSM D.

If the user moves from the landing page of a personality on OSM A to the personality's pool of content in the pollinator, they are faced with a choice set of content rather than a choice set of OSMs. The pools contain the specific pieces of content that personalities want their users to see. Since the user is aware of the curated nature of the content, their engagement is necessarily influenced by the personality's revealed preference for their own content. The pools represent the most pollen-rich environments, offering direct access to the most engaging content from the personality that has captured the user's attention. As a result, the level of engagement with content selected via the personality's pool is likely to be greater than had the user encountered that same piece of content on the personality's landing page.

From the OSM's point of view, the greater engagement from users arriving from a personality's pool on the pollinator translates into more time on the OSM and increased exposure to advertisers.

Consider heterogeneity among users as an example. Suppose that content comes in two forms: long and short. Suppose type A users prefer short-form media, while type B users prefer long-form media. Assume that user types form their preferences through some mechanism of exposure bias. Type A users favour short-form media because they are accustomed to it. As such, there must exist a median media length that simultaneously maximizes engagement for users of both types. This is depicted in Figure 8.

**Figure 8**

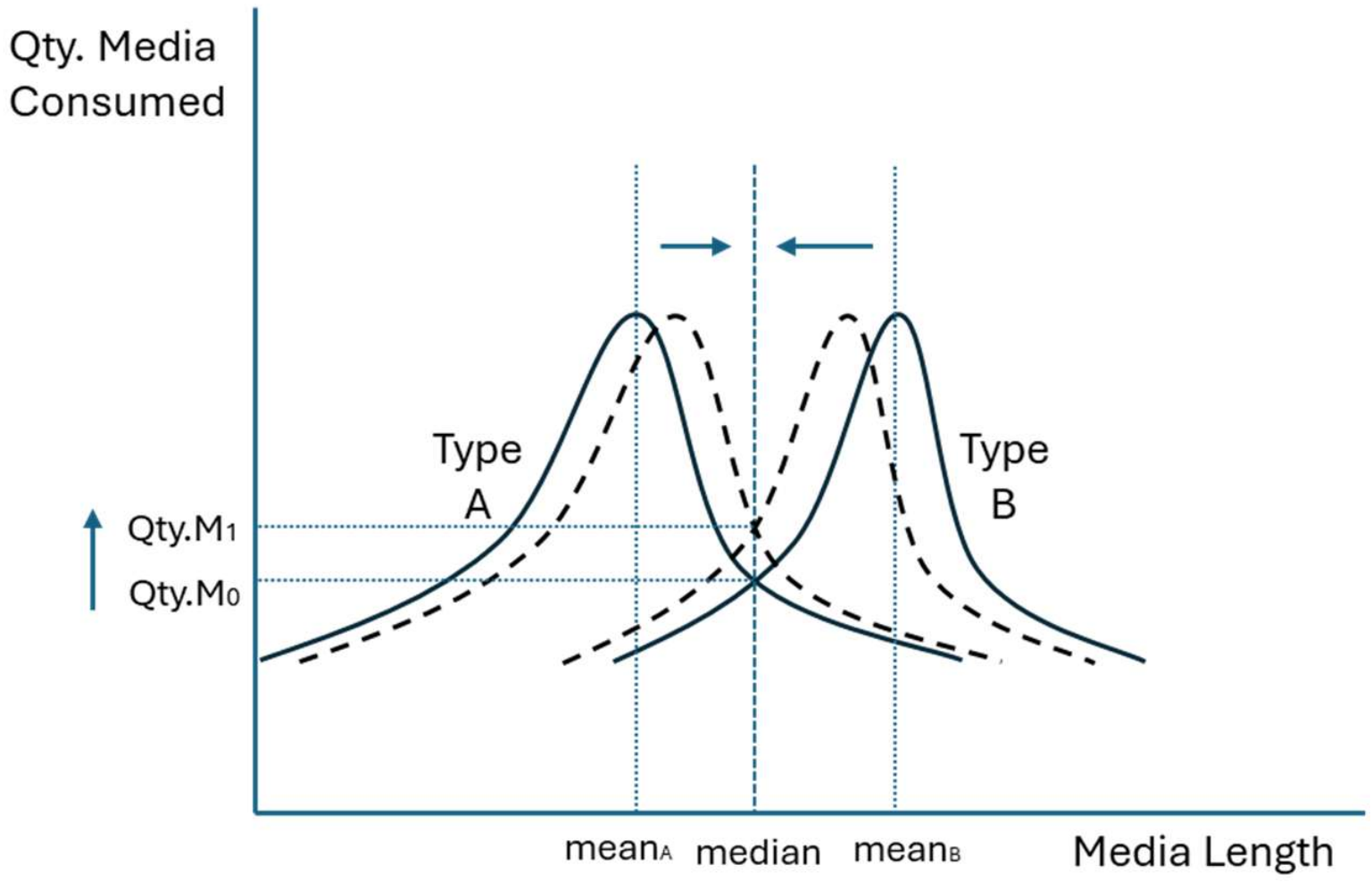

Suppose a type A user is drawn to the curated pool of a personality that produces both types of content. To some degree, type A's bias against long-form media must be tempered by their interest in personality A's content. As more type A users are exposed to long-form media and more type B users are exposed to short-form media, each from their respective preferred personalities, the two user types will naturally become more homogeneous. Greater homogeneity among users implies greater relative capture of the two groups' preference sets by the median media length. If we assume that the median media length reflects media consumption in the most trafficked OSMs, greater capture of user preferences will necessarily benefit the OSMs whose users most closely reflect the median.

## Conclusion

The current web-based social media landscape can be described as fragmented, dominated by a few major platforms amidst a vast array of smaller, niche networks. When users land on a personality's page on any given platform, their engagement tends to be confined to that platform—whether they're exploring content by the initial personality or branching out to others within the same network. While users could theoretically jump between platforms in pursuit of more content, the lack of an immediate and cohesive way to maintain a unified search theme across platforms significantly limits overall traffic to each.

We interpret this search constraint as akin to the natural dynamics of insect-mediated pollen transfer. Just as bees move pollen across various species of plants, users seek out digital content across a dispersed media ecosystem. Each platform is a distinct plant species scattered across fields, with the content acting as the pollen that draws in users. Naturally, platforms with the richest and most diverse content capture the highest user traffic.

Cross-pollination across web-based social media platforms will very likely drives up traffic across the entire ecosystem. Like bees moving from flower to flower, users who can flow seamlessly between platforms increase the volume and diversity of content consumed, prolonging their engagement time.

In this context, we consider the impact of a social media aggregator / pass-through apps such as Link.Tree, Link.Me and &Share that function as pollinators, seamlessly guiding users across pages and platforms in a single browsing session. We also consider "pools" of content: curated collections that showcase a personality's presence across platforms, serving as rich content nodes that encourage deeper, more prolonged engagement compared to standard user interactions.

A cross-platform pollinator app, allowing users to explore content across networks while maintaining their search continuity enhances traffic in proportion to each platform's existing user base. Large platforms capture the largest share of this increase in traffic, engagement duration, and depth. However, due to the diversity of user interests, smaller platforms significantly benefit from cross-pollination having strong incentives to participate, even as it primarily benefits their larger competitors.

This digital synergy, facilitated by a pollinator app capable of aggregating, pooling, and seamlessly distributing interconnected content across the social media ecosystem, represents a positive sum gain for online media. Platforms across the spectrum—from dominant networks to niche communities—stand to benefit from enhanced traffic, deeper engagement, and broader exposure. However, the impact of cross-pollination likely extends far beyond platform-specific metrics, raising compelling questions about broader societal implications.

Future research could explore the transformative potential of cross-pollination for content consumption, user engagement patterns, and information dissemination. As content pools and cross-pollination mechanisms become more sophisticated, they may reshape user experience, tailoring digital environments to individual interests and enhancing personalization in ways that could redefine online interactions. On a larger scale, this phenomenon might foster the development of novel incentive models, where value generation is not confined to individual platforms but distributed across the ecosystem, challenging traditional advertising and revenue strategies.

Moreover, cross-pollination raises important questions about information flow and content diversity, warranting further research on the effects of inter-platform engagement on user perspectives and community building. Studies could explore whether the ability to easily navigate across platforms promotes exposure to diverse viewpoints or reinforces existing echo chambers. The diversity of content recommendation algorithms across different platforms could help break users out of bubbles as echo chambers form along sufficiently different lines on different platforms that pollinators allow for increased information flow from outside of the users' bubble. Alternatively, and somewhat pessimistically, the introduction of the pollinator could blur the boundaries of individual echo chambers creating a larger distributed chamber.

Investigating these questions will be crucial for understanding the full societal implications of an interconnected social media ecosystem and the ethical responsibilities of platforms and developers in shaping such an environment.

Pollinator apps simplify identity resolution and platform navigation across the full stack of digital experience. In doing so, it points to a foundational critique that has gone unaddressed for nearly two decades. The internet is no longer singular, and its search logic is no longer sufficient. Reuniting fragmented identity, discovery, and communication across social media platforms is not an optional luxury. It is a structural necessity for the next stage of the web's evolution.

We envision future studies examining the ways cross-pollination could be leveraged to support educational, civic, professional, medical, property, and legal initiatives, broadening the boundaries of social media's role in modern society. Whether fostering global discourse, advancing access to medical knowledge, streamlining property and real estate networks, facilitating legal information sharing, or creating unified digital communities, the potential applications of content pools and cross-platform dynamics are vast. These explorations will be key to realizing and responsibly managing the far-reaching impact of this emerging digital phenomenon.

## Appendix 1

## Proof

From the definition of $N$,

$$N\sigma = \sum_{k=0}^{\infty} A_k\, Q^k \sigma.$$

Hence

$$G_n(\sigma) = \sum_{i \in S_n} \tau_i\, (N\sigma)_i = \sum_{i \in S_n} \tau_i \sum_{k=0}^{\infty} A_k\, (Q^k \sigma)_i.$$

Because $A_k \geq 0$, $Q^k \geq 0$entrywise, and $\tau_i \geq 0$, we may rearrange terms:

$$G_n(\sigma) = \sum_{k=0}^{\infty} A_k \sum_{i \in S_n} \tau_i\, (Q^k \sigma)_i.$$

Write the $k$-step contribution as

$$g_n^{(k)}(\sigma) \equiv \sum_{i \in S_n} \tau_i\, (Q^k \sigma)_i.$$

It is therefore enough to show that for every $k \geq 0$,

$$g_n^{(k)}(\sigma) \geq g_m^{(k)}(\sigma) \text{whenever } s_n \geq s_m.$$

### Step 1: The case $k = 0$

Since $Q^0 = I$,

$$g_n^{(0)}(\sigma) = \sum_{i \in S_n} \tau_i\, \sigma_i.$$

By entry monotonicity, the total initial mass on platform $n$is weakly increasing in $s_n$. By A3, the average time spent per visited state on platform $n$is also weakly increasing in $s_n$.

Therefore, the initial-period attention contribution satisfies

$$g_n^{(0)}(\sigma) \geq g_m^{(0)}(\sigma).$$

**Step 2: The case $k \geq 1$**

For any $k \geq 1$,

$$(Q^k \sigma)_i = \sum_{j \in S} (Q^{k-1} \sigma)_j q_{ij}$$

Substituting into $g_n^{(k)}(\sigma)$,

$$g_n^{(k)}(\sigma) = \sum_{i \in S_n} \tau_i \sum_{j \in S} (Q^{k-1} \sigma)_j q_{ij} = \sum_{j \in S} (Q^{k-1} \sigma)_j \sum_{i \in S_n} \tau_i \, q_{ij}.$$

Since average per-state attention on platform $n$is weakly increasing in size, the inner term is bounded below by

$$\sum_{i \in S_n} \tau_i \, q_{ij} \;\; \geq \;\; \underline{\tau}_n \sum_{i \in S_n} q_{ij} \, ,$$

where $\underline{\tau}_n$denotes any lower bound consistent with platform-$n$'s average retention, and similarly for platform $m$. By transition monotonicity, for every origin state $j$, total transition probability into platform $n$is weakly increasing in $s_n$. Hence, when $s_n \geq s_m$,

$$\sum_{i \in S_n} q_{ij} \;\; \geq \;\; \sum_{i \in S_m} q_{ij} \; \forall j \in S.$$

Combining with $\overline{\tau}_n = \frac{1}{|S_n|} \sum_{i \in S_n} \tau_i$ yields

$$\sum_{i \in S_n} \tau_i \, q_{ij} \;\geq\; \sum_{i \in S_m} \tau_i \, q_{ij} \forall j \in S.$$

Since $(Q^{k-1}\sigma)_j \geq 0$ for all $j$, summing over $j$gives

$$g_n^{(k)}(\sigma) \geq g_m^{(k)}(\sigma).$$

**Step 3: Sum across all stages**

Because $A_k \geq 0$for all $k$,

$$G_n(\sigma) - G_m(\sigma) = \sum_{k=0}^{\infty} A_k \left(g_n^{(k)}(\sigma) - g_m^{(k)}(\sigma)\right) \geq 0.$$

Thus

$$G_n(\sigma) \geq G_m(\sigma).$$

If monotonicity is strict on a set of positive measure, then at least one stage-$k$ contribution is strictly larger for platform $n$, and therefore

$$G_n(\sigma) > G_m(\sigma).$$

This proves the claim.